\documentclass[prl,twocolumn,showpacs,superscriptaddress,preprintnumbers,amssymb]{revtex4}
\usepackage{graphicx}
\usepackage{dcolumn}
\usepackage{bm}
\usepackage{wasysym}

\newcommand{\beq}{\begin{equation}}
\newcommand{\eeq}{\end{equation}}
\newcommand{\beqn}{\begin{eqnarray}}
\newcommand{\eeqn}{\end{eqnarray}}

\begin{document}
\title{Time-Reversal Phase Transition at the Edge of 3d Topological Band Insulator}
\author{Cenke Xu}
\affiliation{Department of Physics, Harvard University, Cambridge,
MA 02138}
\date{\today}

\begin{abstract}

We study the time reversal (T) symmetry breaking of 2d helical
fermi liquid, with application to the edge states of 3d
topological band insulators with only one two-component Dirac
fermion at finite chemical potential, as well as other systems
with spin-orbit coupling. The T-breaking Ising order parameter is
not over-damped and the theory is different from the ordinary
Hertz-Millis theory for order parameters at zero momentum. We
argue that the T-breaking phase transition is an 3d Ising
transition, and the quasiparticles are well-defined in the quantum
critical regime.


\end{abstract}
\pacs{} \maketitle

Time reversal (T) symmetry is the key to guarantee the stability
of both 2d and 3d topological insulators (TBI)
\cite{kane2005a,kane2005b,fu2007,fu2007a}, therefore it is
meaningful to study the T-symmetry breaking in these systems.
Because the bulk of TBI is always an insulator, the T-breaking
transition only involves the edge states, which are gapless in a
T-symmetric phase, and the spectrum opens up a gap when T is
broken. The simplest version of 3d TBI has only one two-component
Dirac fermion at the edge, which can be perfectly realized in
materials based on $\mathrm{Bi}_{2} \mathrm{Se}_3$ and
$\mathrm{Bi}_{2} \mathrm{Te}_3$
\cite{hasan2009a,hasan2009b,fang2008,zxshen2009}. The time
reversal symmetry can either be broken explicitly by magnetic
impurities, or broken spontaneously by strong enough interactions.
The effects of magnetic impurities and quenched disorders on the
edge states of 2d and 3d TBI has been discussed in Ref.
\cite{xuedge,wuedge} and Ref. \cite{xuzhang} respectively.
Spontaneous T-breaking phase transition is most relevant to the
transition metal version of the 3d TBI with interplay between
spin-orbit coupling and strong interaction \cite{balents2009}, and
it is the goal of the current paper.

Without loss of generality, the edge state of 3d TBI is described
by the following time-reversal invariant Lagrangian
\cite{fu2007,fu2008}: \beqn L_f = \bar{\psi}(\gamma_0 (i\partial_t
- \mu) + v_f i \gamma_j\partial_j)\psi. \label{dirac} \eeqn
$\gamma^0 = \sigma^z$, $\gamma^1 = i\sigma^x$, $\gamma^2 =
i\sigma^y$, $\bar{\psi} = \psi^\dagger \gamma^0$. $v_f$ is the
fermi velocity at the Dirac point, $\mu$ is the chemical
potential. The Pauli matrices in Eq. \ref{dirac} represent the
pseudospin, which is a combination between real spin space and
orbital space. For conciseness we will call $\sigma^a$ the spin
hereafter. The spin $\sigma^a$ of the electrons are perpendicular
with their momenta. This helical spin alignment has been
successfully observed in a recent ARPES measurement
\cite{hasan2009}. The T-symmetry guarantees that in the Lagrangian
the Dirac mass gap $\bar{\psi}\psi$ does not appear explicitly,
although a mass generation can occur when the T-symmetry is
spontaneously broken. The Dirac gap is simply the $z-$spin
magnetization, hence the gap can be spontaneously generated with
strong enough ferromagnetic interaction between $z-$component of
spins: $-(\bar{\psi}\psi)_{r^\prime}V_{\vec{r},
\vec{r}^\prime}(\bar{\psi}\psi)_{\vec{r}^\prime}$. To describe
this T-breaking transition, we can define an Ising order parameter
$\phi$, which couples to the Dirac fermions as \beqn L &=& L_f +
L_b + L_{bf}, \cr\cr L_b &=& (\partial_t \phi)^2 - \sum_{i = x,y}
v_b^2 (\partial_i\phi)^2 - r \phi^2 - u\phi^4, \cr L_{bf} &=& g
\phi \bar{\psi}\psi. \label{lag}\eeqn $\bar{\psi}\psi$ order
breaks T, and drives the edge to a quantum Hall phase. Identifying
the leading spin order instability requires detailed knowledge of
the fermion interaction, hence we focus on the universal physics
at the quantum critical point, assuming the existence of the phase
transition. In the current work we only discuss the discrete
symmetry breaking, the transition with continuous symmetry
breaking will be studied in another paper \cite{xucontinuous}. The
Lagrangian Eq. \ref{lag} can also describe the phase transition of
magnetic impurities doped into the system, and the order parameter
$\phi$ stands for the global magnetization of the magnetic
impurities. The $u\phi^4$ term represents either the
self-interaction between the magnetic impurities, or the higher
order spin-spin interactions between helical fermions. In this
paper we assume $u
> 0$ and large enough to ensure a second order transition.

Let us first take $\mu = 0$ in Eq. \ref{lag}, now this model
becomes the Higgs-Yukawa model, which is believed to be equivalent
to the Gross-Neveu model \cite{gross1974,wilson1973} $L =
i\bar{\psi}\gamma_\mu\partial_\mu\psi - \gamma(\bar{\psi}\psi)^2,
\label{gn}$ at least when $v_f = v_b$. The transition of $\phi$ is
not 3d Ising transition because the coupling $g$ is relevant at
the 3d Ising fixed point, based on the well-known scaling
dimensions $[\psi] = 1/2$, and $[\phi] = (d - 1)/2 + \eta/2 =
0.518$ at the 3d Ising fixed point \cite{hasenbusch1998}. If there
are $N$ flavors of Dirac fermions, The critical exponents of this
transition with large $N$ have been calculated by means of $1/N$
and $\epsilon = 4 - d$ expansions
\cite{zustin1991,petersson1994,gracey1991,gracey1992}, and a
second order transition with non-Ising universality class was
found. In our current case with $N = 1$, there is no obvious small
parameter to expand, we conjecture that the transition is still
second order, with different universality class from the 3d Ising
transition.

Let us now turn on a finite chemical potential $\mu$, but still
make $\mu$ much smaller than the band-width $2 \Lambda$ of the
edge states. Now the edge states become a helical fermi liquid,
with spins aligned parallel with its fermi surface. The tuning
parameter $r$ in Eq. \ref{lag} will be renormalized by the static
and uniform susceptibility of $\sigma^z$ of the helical fermi
liquid \beqn \Delta r \phi^2 = \mathrm{Re}[\chi(0,0)]\phi^2 \sim
g^2(\mu - \Lambda) \phi^2. \eeqn Therefore the phase transition of
$\phi$ can be driven by tuning the chemical potential $\mu$. Also,
it is straightforward though a little tedious to check that the
momentum and frequency dependence of $\mathrm{Re}[\chi]$ are
nonsingular: $\mathrm{Re}[\chi(\omega, q)] \sim c_0 - c_1\omega^2
- c_2q^2 + \cdots$.

\begin{figure}
\includegraphics[width=2.5in]{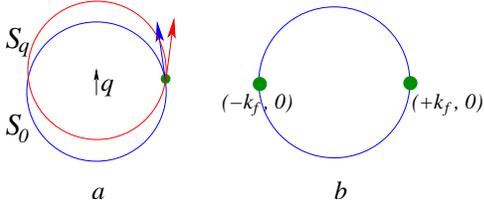}
\caption{The fermi surface of Dirac fermions, with finite chemical
potential. $a$, when we translate the fermi surface with a small
momentum $\vec{q}$, at the intersection the spins are almost
parallel; $b$, the two patches of fermi surface Eq. \ref{lag2}
describes.} \label{fs}
\end{figure}

\begin{figure}
\includegraphics[width=3.1in]{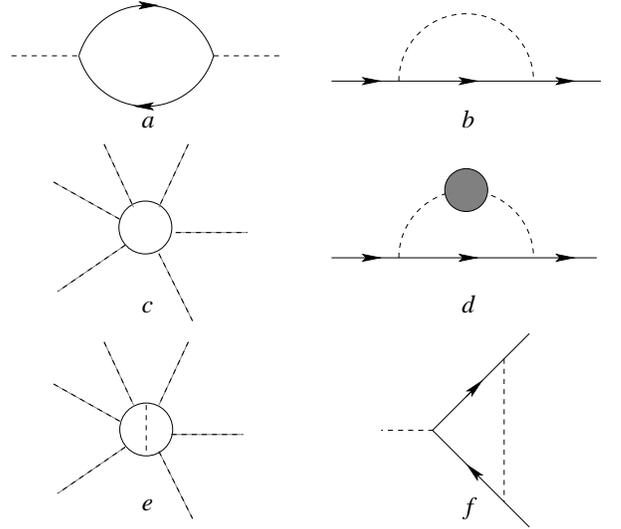}
\caption{The one-loop Feynman diagrams for boson, fermion
self-energy, vertex correction, and $\phi^n$ term generated with
fermion loop. The dashed line and solid line represent the $\phi$
propagator and fermion propagator respectively.} \label{fd}
\end{figure}

As the ordinary Hert-Millis theory \cite{hertz1976} of quantum
phase transition inside fermi liquid, the singular correction to
the effective Lagrangian of order parameter comes from the
imaginary part of the susceptibility. At the critical point, the
critical mode of Ising order parameter $\phi$ can be damped
through particle-hole excitations. The damping rate can be
calculated from the Feynman diagram Fig. \ref{fd}$a$, or through
the Fermi-Golden rule \beqn \mathrm{Im}[\Sigma_{\phi}(\omega, q)]
&\sim& \int \frac{d^2k}{(2\pi)^2} [f(\epsilon_{k+q}) -
f(\epsilon_{k})]\cr\cr &\times& \delta(|\omega| - \epsilon_{k+q} +
\epsilon_k)|\langle k| g \bar{\psi}_k\psi_{k+q} |k+q \rangle|^2
\cr\cr &\sim& \ g^2 \frac{|\omega| q}{ v_fk_f^2} \sqrt{1-
\frac{\omega^2}{v_f^2q^2}}. \label{damp}\eeqn This result is
obtained in the limit $q \ll k_f$, $k_f$ is the fermi wave-vector.
When $|\omega| > v_fq$ the scattering rate vanishes for kinematic
reasons, therefore when $v_b > v_f$ this decay rate is unimportant
because the Green's function of $\phi$ will peak when $\omega \sim
v_bq$. From now on we will assume that $v_b < v_f$. The decay rate
obtained above differs from the Hertz-Millis theory
\cite{hertz1976} which usually takes the form $|\omega|/q$ for
order parameters at zero momentum. This result can be physically
understood as following: $\phi(\vec{q})$ can transfer momentum
$\vec{q}$ to the fermi surface, and if we denote the fermi surface
as $\mathcal{S}_0$, and denote the fermi surface translated by a
small momentum $\vec{q}$ as $\mathcal{S}_{\vec{q}}$, then as long
as $\vec{q}$ is small enough $\mathcal{S}_{\vec{q}}$ and
$\mathcal{S}_0$ will have almost the same spin directions at their
intersection. Because $\sigma^z$ always flips spin direction in
the XY plane, when two spins are parallel the matrix element of
$\sigma^z$ vanishes. Mathematically this intuition is manifested
as $|\langle k| \bar{\psi}_k\psi_{k+q} |k+q \rangle|^2$ vanishes
as $q^2$ in the limit of $q \rightarrow 0$. Therefore in this case
$\phi$ is not overdamped at low momentum and frequency.

If we ignore the self-interaction between $\phi$, and take the
Gaussian part of $L_b$, we can calculate the self-energy
correction of fermion $\psi$ through Feynman diagram Fig.
\ref{fd}$b$. Evaluated close to $|\nu| \sim \epsilon_{q}$, the
imaginary part of fermion self-energy scales as \beqn
\Sigma(\nu)'' &\sim& \int d^2k \frac{1}{\omega_k}
[\theta(\epsilon_{k+q}) \delta(\nu - \epsilon_{k+q} -
\omega_k)\cr\cr &-& \theta(- \epsilon_{k+q}) \delta(\nu -
\epsilon_{k+q} + \omega_k)] \cr\cr &\times& |\langle q|
g\bar{\psi}_q\psi_{k+q} |k+q \rangle|^2 \sim g^2
\nu^2\mathrm{Sign}[\nu] + \cdots \label{psiself}\eeqn Unlike the
Hertz-Millis theory, the scaling of $\Sigma(\nu)''$ is similar to
fermi liquid, which means that the quasiparticles are well-defined
even at the quantum critical point.

The above calculations are only one-loop level. To evaluate higher
loop diagrams, we had better simplify the problem by considering
two patches of the fermi surface around two opposite points $(\pm
k_f , 0)$, and label the fermions in terms of its momentum $p_x =
k_x - k_f$, $p_y = k_y$. Now the action becomes \beqn L_f &=&
\psi^\dagger(\vec{p}) (\omega - v_f p_x \tau^z - v_y
p^2_y)\psi(\vec{p}), \cr\cr L_b &=& \eta \omega^2
|\phi(\vec{p})|^2 - (v^2_{bx}p^2_x +
v_{by}^2p_y^2)|\phi(\vec{p})|^2 + \cdots \cr\cr L_{bf} &=& i
gq_y\phi(\vec{q}) \psi^\dagger(\vec{p})\tau^z
\psi(\vec{p}+\vec{q})+ \cdots \label{lag2}\eeqn Here both $|p_x|$
and $|p_y|$ are much smaller than $k_f$, and $v_y = v_f/(2k_f)$.
$\tau^z$ is the Pauli matrix operating on the space of two fermi
patches $(\pm k_f, 0)$.
This isolated patch approximation is based on the observation that
$\phi_{\vec{q}}$ most strongly couples to the patch with
$\vec{q}\perp \vec{K}_f$, where the particle-hole excitation with
momentum $\vec{q}$ is soft. Also, at low energy limit, none of the
scattering process will mix these fermions with those from other
patches. For instance, if we integrate out the boson
$\phi_{\vec{q}}$, interaction between different patches will be
induced, but the standard scaling argument for ordinary fermi
liquid suggests that the only important interaction at low energy
has $\vec{q} = 0$ $i.e.$ the $\delta n_\theta \delta
n_{\theta^\prime}$ interaction. However, when $\vec{q} = 0$ the
interaction vertex vanishes. Therefore the isolated patch
approximation is reasonable.

Under discrete symmetry transformations $\mathrm{T}$,
$\mathrm{P}_x$ and $\mathrm{P}_y$, the physical quantities in Eq.
\ref{lag2} transform as \beqn \mathrm{T}&:& \ t \rightarrow -t, \
\psi \rightarrow \tau^x \psi, \ k_i \rightarrow - k_i, \ \phi
\rightarrow -\phi, \ i \rightarrow -i, \cr \mathrm{P}_x &:& \ x
\rightarrow -x, \ \psi \rightarrow \tau^x \psi, \ k_x \rightarrow
- k_x, \ \phi \rightarrow -\phi ,\cr \mathrm{P}_y &:& \ y
\rightarrow -y, \ \psi \rightarrow \psi, \ k_y \rightarrow -k_y, \
\phi \rightarrow -\phi. \label{symmetry}\eeqn and the action is
invariant. Had we only kept one single fermi patch at $(+k_f, 0)$
like Ref. \cite{polchinksi1994,sslee2009}, the action would not be
invariant under these discrete transformations.

The fermion-boson vertex is proportional to $q_y$ of $\phi$,
therefore for any loop diagram with $\phi$ external line, the loop
diagram will vanish as $q_y \rightarrow 0$ for each $\phi$
external line. There are two different ways to assign scaling
dimensions to operators in Eq. \ref{lag2}: \beqn \mathrm{Scaling}
&& \mathbf{1}, \ [\omega] = 1, \ [p_x] = 1, \ [p_y] = 1, \ [v_y] =
-1, \cr [\phi] &=& - \frac{5}{2}, \ [\psi] = -2, \ [\eta] =
[v_{bx}] = [v_{by}] = 0, \cr [g] &=& -\frac{1}{2}, \cr\cr
\mathrm{Scaling}&& \mathbf{2}, \ [\omega] = 2, \ [p_x] = 2, \
[p_y] = 1, \ [v_y] = 0, \cr [\phi] &=& [\psi] = - \frac{7}{2}, \
[\eta] = [v_{bx}^2] = -2, \ [v_{by}^2] = 0, \cr [g] &=&
-\frac{1}{2}. \label{scaling1}\eeqn For both scaling choices, $[g]
< 0$, $i.e.$ according to the naive scaling the coupling between
fermions and bosons are irrelevant, and the loop diagrams are
suppressed. When we evaluate loop integrals, irrelevant terms can
in general be ignored, but in order to avoid divergence from
integrating a constant, we have to make a diagram-dependent choice
of scaling from the two options in Eq. \ref{scaling1}, otherwise
some irrelevant terms have to be kept in the integral. For
instance, we can reproduce the results obtained previously from
scaling argument: at the $g^2$ order, choosing the second scaling
in Eq. \ref{scaling1}, the self-energy correction of $\phi$ should
have dimension 3, which is consistent with the direct calculation
with action Eq. \ref{lag2} and Feynman diagram Fig. \ref{fd}$a$:
\beqn \mathrm{Im}[\Sigma_{\phi}] \sim g^2 |\omega||q_y|,
\label{aniself}\eeqn which due to energy conservation is valid
when $|\omega - v_f q_x| < v_f|q_y|$.
For the fermion
self-energy, in order to avoid naive divergence one has to choose
the first set of scaling dimensions, $[g^2] = -1$ implies that the
self-energy should have dimension 2, which is consistent with the
result $\Sigma(\nu)'' \sim g^2 \nu^2$ we obtained before. The one
loop vertex correction can be calculated using the second scaling
and Fig. \ref{fd}$f$, the result is $V_q \sim q_y^2/(|q_y| + c g^2
|\omega|)$.

Now let us discuss the nature of the T-breaking transition. The
pure boson Lagrangian $L_b$ in Eq. \ref{lag} describes a 3d Ising
transition. At the $g^2$ order the perturbation at the 3d Ising
transition is included in the self-energy correction to $\phi$,
whose singular contribution is in the imaginary part.
The imaginary part of the self-energy is given by both Eq.
\ref{damp} and Eq. \ref{aniself}, evaluated with the the full
fermi surface and isolated patch approximation respectively. In
both cases this self-energy mix $\phi$ at distinct points in
space-time, their actual scaling dimensions at the 3D Ising
critical point can be estimated as $ D - (2 + D - 2 +\eta) = -
\eta$, $\eta \sim 0.037$ \cite{hasenbusch1998}. Therefore at the
$g^2$ order there is no relevant perturbation induced at the 3d
Ising fixed point.

The higher loop diagrams are more complicated, although in the
previous paragraph we showed that in both choices of scalings $g$
is irrelevant, it does not immediately imply none of the higher
order loops can generate important terms at the 3d Ising fixed
point. This is because when we evaluate the fermi loop, in order
to avoid naive divergence we have to take the second scaling in
Eq. \ref{scaling1}, which is different from the 3d Ising fixed
point with isotropic scaling dimensions in space-time. For
instance the leading $\phi^n$ term generated at $g^n$ order
perturbation is given by diagram Fig. \ref{fd}$c$, which should
take the form \beqn g^n \prod_{i = 1}^n
[q_{i,y}\phi(\vec{q}_i)]\times f_n(\omega_j, \vec{q}_j). \eeqn
Notice that all the $\phi^n$ terms with $n$ odd are forbidden by
symmetry. This term is irrelevant based on the second scaling of
Eq. \ref{scaling1}, but in order to know its scaling dimension at
the 3d Ising fixed point, we need to evaluate its form more
explicitly. The function $f(\omega_j, \vec{q}_j)$ is integral of
the following fermion loop: \beqn f_n(\omega_j, \vec{q}_j) \sim
\int d\omega dp_x dp_y \times \delta(\sum
\omega_j)\delta(\sum\vec{q}_j)\cr \times \mathrm{Tr}[\prod_{j =
1}^n G(\omega + \sum_{i = 1}^j\omega_i, \vec{p} + \sum_{i = 1}^j
\vec{q}_{i})]. \eeqn After the integral, this term has a very
complicated dependence of the external frequency $\omega_j$ and
momentum $\vec{p}_j$, but since we are only interested in its
scaling dimension, the following schematic form will be good
enough: \beqn f_n \sim \sum\frac{|\Omega|}{\sum |Q_{y}|\prod_{j =
1}^{n-2}(\Omega_{j} + v_fQ_{jx}) + \cdots}. \label{phin} \eeqn
$\Omega$ and $Q$ represent linear combination between external
frequency and momentum $q$ respectively. In the denominator, the
ellipses include terms with higher power of momentum compared with
the leading term. We can easily verify that when $n = 2$ Eq.
\ref{phin} reproduces the well-known result $|\omega|/|q_y|$. At
the 3d Gaussian fixed point, the coefficient of the $\phi^n$ term
will have scaling dimension $ 1 - n/2$, which should be irrelevant
for any $n \geq 4$. Eq. \ref{phin} is applicable to the kinematic
regime with all the external momenta nearly parallel to $\hat{y}$,
when $\phi$ couples most strongly with particle-hole excitations.
For more general kinematic regime the $\phi^n$ term generated is
expected to be no more singular than Eq. \ref{phin}.

So far we have only considered the leading $\phi^n$ term, which is
generated at $g^n$ order. Higher order contribution to $\phi^n$
always involve one or more internal boson lines like Fig.
\ref{fd}$e$, and because of the suppression of $p_y$ at the
internal vertices, we expect these higher order terms will not be
more relevant than the leading order. For instance the result of
diagram Fig. \ref{fd}$e$ with one internal boson line has the same
scaling dimension as Fig. \ref{fd}$c$. Based on these
observations, the T-breaking phase transition in the helical fermi
liquid with finite $\mu$ is expected to be a 3d Ising transition.
If we take into account of the interaction between $\phi$ at the
3d Ising universality class $i.e.$ using the fully dressed boson
propagator in Fig. \ref{fd}$d$, the self-energy of the fermion
will be even more suppressed due to self-screening between bosons.
One reasonable result could be $\Sigma(\nu)'' \sim g^2 |\nu|^{2 +
\eta}\mathrm{Sign[\nu]}, $ $\eta \sim 0.037$ \cite{hasenbusch1998}
is the anomalous dimension of $\phi$ at the 3d Ising transition,
since $\eta > 0$, the quasiparticle is always well-defined at the
quantum critical regime.

The Lagrangian Eq. \ref{dirac} is invariant when spin and space
are rotated by the same and arbitrary angle, which is generically
larger than the symmetry of the microscopic system. For instance
in material $\mathrm{Bi}_{2 - x}\mathrm{Sn}_x \mathrm{Te}_3$ the
fermi surface of edge states is not circular when the chemical
potential is large, instead it is a hexagonal star with six sharp
corners \cite{zxshen2009}. Therefore with large chemical
potential, terms with higher order momentum should be considered
in the free electron Lagrangian $L_f$ of Eq. \ref{dirac}. These
higher order terms can lead to many new effects, for instance it
may align the spins slightly along $\hat{z}$ direction instead of
completely within the XY plane \cite{liu2009,fu2009}, although the
integral of $\sigma^z$ vanishes along the whole fermi surface. If
the spins have $\hat{z}$ component, then $\phi \sim
\bar{\psi}\psi$ will cause a deformation of the fermi surface, and
is overdamped for small momentum, in this case the ordinary $z =
3$ Hertz-Millis theory becomes applicable.

In summary, we studied the time-reversal symmetry breaking for
single Dirac fermion with finite chemical potential. Unlike the
ordinary Hertz-Millis theory, the Ising order parameter is not
overdamped, and we argue that the coupling between Ising order
parameter and fermions is weak in the infrared limit. The
transition most likely belongs to the 3d Ising universality class.
The analysis in our paper can be generalized to many other
systems. For instance we can consider the spin order $\psi^\dagger
\sigma^z \psi$ in the Rashba model \cite{rashba1960,rashba1984}
with inner and outer fermi surfaces with opposite inplane helical
spin direction, and the results are very similar to our paper.
Another system is graphene with $N = 4$ flavors of Dirac fermion,
our analysis applies to order parameters $\bar{\psi}\psi$ and
$\bar{\psi}T^a\psi$ ($T^a \in \mathrm{SU}(N)$). For instance the
phase transition of Quantum Spin Hall order
$\bar{\psi}\vec{S}\psi$ belongs to the 3d O(3) universality class,
when the fermi energy is tuned away from the Dirac point. In
future we shall try to make connection between our results and
realistic physical system, after a suitable physical system with
both topological band structure and strong interaction is
discovered, like the one studied theoretically in Ref.
\cite{balents2009}.

The author appreciate the very helpful discussion with Max
Metlitski and Xiaoliang Qi. This work is sponsored by the Society
of Fellows, Harvard University.



\bibliography{domain}

\begin{thebibliography}{29}
\expandafter\ifx\csname natexlab\endcsname\relax\def\natexlab#1{#1}\fi
\expandafter\ifx\csname bibnamefont\endcsname\relax
  \def\bibnamefont#1{#1}\fi
\expandafter\ifx\csname bibfnamefont\endcsname\relax
  \def\bibfnamefont#1{#1}\fi
\expandafter\ifx\csname citenamefont\endcsname\relax
  \def\citenamefont#1{#1}\fi
\expandafter\ifx\csname url\endcsname\relax
  \def\url#1{\texttt{#1}}\fi
\expandafter\ifx\csname urlprefix\endcsname\relax\def\urlprefix{URL }\fi
\providecommand{\bibinfo}[2]{#2}
\providecommand{\eprint}[2][]{\url{#2}}

\bibitem[{\citenamefont{Kane and Mele}(2005{\natexlab{a}})}]{kane2005a}
\bibinfo{author}{\bibfnamefont{C.~L.} \bibnamefont{Kane}} \bibnamefont{and}
  \bibinfo{author}{\bibfnamefont{E.~J.} \bibnamefont{Mele}},
  \bibinfo{journal}{Phys. Rev. Lett} \textbf{\bibinfo{volume}{95}},
  \bibinfo{pages}{226801} (\bibinfo{year}{2005}{\natexlab{a}}).

\bibitem[{\citenamefont{Kane and Mele}(2005{\natexlab{b}})}]{kane2005b}
\bibinfo{author}{\bibfnamefont{C.~L.} \bibnamefont{Kane}} \bibnamefont{and}
  \bibinfo{author}{\bibfnamefont{E.~J.} \bibnamefont{Mele}},
  \bibinfo{journal}{Phys. Rev. Lett} \textbf{\bibinfo{volume}{95}},
  \bibinfo{pages}{146802} (\bibinfo{year}{2005}{\natexlab{b}}).

\bibitem[{\citenamefont{Fu et~al.}(2007)\citenamefont{Fu, Kane, and
  Mele}}]{fu2007}
\bibinfo{author}{\bibfnamefont{L.}~\bibnamefont{Fu}},
  \bibinfo{author}{\bibfnamefont{C.~L.} \bibnamefont{Kane}}, \bibnamefont{and}
  \bibinfo{author}{\bibfnamefont{E.~J.} \bibnamefont{Mele}},
  \bibinfo{journal}{Phys. Rev. Lett.} \textbf{\bibinfo{volume}{98}},
  \bibinfo{pages}{106803} (\bibinfo{year}{2007}).

\bibitem[{\citenamefont{Fu and Kane}(2007)}]{fu2007a}
\bibinfo{author}{\bibfnamefont{L.}~\bibnamefont{Fu}} \bibnamefont{and}
  \bibinfo{author}{\bibfnamefont{C.~L.} \bibnamefont{Kane}},
  \bibinfo{journal}{Phys. Rev. B} \textbf{\bibinfo{volume}{76}},
  \bibinfo{pages}{045302} (\bibinfo{year}{2007}).

\bibitem[{\citenamefont{Xia et~al.}(2009)\citenamefont{Xia, Qian, Hsieh, Wray,
  Pal, Lin, Bansil, Grauer, Hor, Cava et~al.}}]{hasan2009a}
\bibinfo{author}{\bibfnamefont{Y.}~\bibnamefont{Xia}},
  \bibinfo{author}{\bibfnamefont{D.}~\bibnamefont{Qian}},
  \bibinfo{author}{\bibfnamefont{D.}~\bibnamefont{Hsieh}},
  \bibinfo{author}{\bibfnamefont{L.}~\bibnamefont{Wray}},
  \bibinfo{author}{\bibfnamefont{A.}~\bibnamefont{Pal}},
  \bibinfo{author}{\bibfnamefont{H.}~\bibnamefont{Lin}},
  \bibinfo{author}{\bibfnamefont{A.}~\bibnamefont{Bansil}},
  \bibinfo{author}{\bibfnamefont{D.}~\bibnamefont{Grauer}},
  \bibinfo{author}{\bibfnamefont{Y.~S.} \bibnamefont{Hor}},
  \bibinfo{author}{\bibfnamefont{R.~J.} \bibnamefont{Cava}},
  \bibnamefont{et~al.}, \bibinfo{journal}{Nature Physics}
  \textbf{\bibinfo{volume}{5}}, \bibinfo{pages}{398} (\bibinfo{year}{2009}).

\bibitem[{\citenamefont{Hsieh et~al.}(2009{\natexlab{a}})\citenamefont{Hsieh,
  Xia, Qian, Wray, Dil, Meier, Osterwalder, Patthey, Checkelsky, Ong
  et~al.}}]{hasan2009b}
\bibinfo{author}{\bibfnamefont{D.}~\bibnamefont{Hsieh}},
  \bibinfo{author}{\bibfnamefont{Y.}~\bibnamefont{Xia}},
  \bibinfo{author}{\bibfnamefont{D.}~\bibnamefont{Qian}},
  \bibinfo{author}{\bibfnamefont{L.}~\bibnamefont{Wray}},
  \bibinfo{author}{\bibfnamefont{J.~H.} \bibnamefont{Dil}},
  \bibinfo{author}{\bibfnamefont{F.}~\bibnamefont{Meier}},
  \bibinfo{author}{\bibfnamefont{J.}~\bibnamefont{Osterwalder}},
  \bibinfo{author}{\bibfnamefont{L.}~\bibnamefont{Patthey}},
  \bibinfo{author}{\bibfnamefont{J.~G.} \bibnamefont{Checkelsky}},
  \bibinfo{author}{\bibfnamefont{N.~P.} \bibnamefont{Ong}},
  \bibnamefont{et~al.}, \bibinfo{journal}{Nature}
  \textbf{\bibinfo{volume}{460}}, \bibinfo{pages}{1101}
  (\bibinfo{year}{2009}{\natexlab{a}}).

\bibitem[{\citenamefont{Zhang et~al.}(2009{\natexlab{a}})\citenamefont{Zhang,
  Liu, Qi, Dai, Fang, and Zhang}}]{fang2008}
\bibinfo{author}{\bibfnamefont{H.}~\bibnamefont{Zhang}},
  \bibinfo{author}{\bibfnamefont{C.-X.} \bibnamefont{Liu}},
  \bibinfo{author}{\bibfnamefont{X.-L.} \bibnamefont{Qi}},
  \bibinfo{author}{\bibfnamefont{X.}~\bibnamefont{Dai}},
  \bibinfo{author}{\bibfnamefont{Z.}~\bibnamefont{Fang}}, \bibnamefont{and}
  \bibinfo{author}{\bibfnamefont{S.-C.} \bibnamefont{Zhang}},
  \bibinfo{journal}{Nature Phys.} \textbf{\bibinfo{volume}{5}},
  \bibinfo{pages}{438} (\bibinfo{year}{2009}{\natexlab{a}}).

\bibitem[{\citenamefont{Chen et~al.}(2009)\citenamefont{Chen, Analytis, Chu,
  Liu, Mo, Qi, Zhang, Lu, Dai, Fang et~al.}}]{zxshen2009}
\bibinfo{author}{\bibfnamefont{Y.~L.} \bibnamefont{Chen}},
  \bibinfo{author}{\bibfnamefont{J.~G.} \bibnamefont{Analytis}},
  \bibinfo{author}{\bibfnamefont{J.~H.} \bibnamefont{Chu}},
  \bibinfo{author}{\bibfnamefont{Z.~K.} \bibnamefont{Liu}},
  \bibinfo{author}{\bibfnamefont{S.~K.} \bibnamefont{Mo}},
  \bibinfo{author}{\bibfnamefont{X.~L.} \bibnamefont{Qi}},
  \bibinfo{author}{\bibfnamefont{H.~J.} \bibnamefont{Zhang}},
  \bibinfo{author}{\bibfnamefont{D.~H.} \bibnamefont{Lu}},
  \bibinfo{author}{\bibfnamefont{X.}~\bibnamefont{Dai}},
  \bibinfo{author}{\bibfnamefont{Z.}~\bibnamefont{Fang}}, \bibnamefont{et~al.},
  \bibinfo{journal}{arXiv:0904.1829}  (\bibinfo{year}{2009}).

\bibitem[{\citenamefont{Xu and Moore}(2006)}]{xuedge}
\bibinfo{author}{\bibfnamefont{C.}~\bibnamefont{Xu}} \bibnamefont{and}
  \bibinfo{author}{\bibfnamefont{J.~E.} \bibnamefont{Moore}},
  \bibinfo{journal}{Phys. Rev. B} \textbf{\bibinfo{volume}{73}},
  \bibinfo{pages}{045322} (\bibinfo{year}{2006}).

\bibitem[{\citenamefont{Wu et~al.}(2006)\citenamefont{Wu, Bernevig, and
  Zhang}}]{wuedge}
\bibinfo{author}{\bibfnamefont{C.}~\bibnamefont{Wu}},
  \bibinfo{author}{\bibfnamefont{B.~A.} \bibnamefont{Bernevig}},
  \bibnamefont{and} \bibinfo{author}{\bibfnamefont{S.-C.} \bibnamefont{Zhang}},
  \bibinfo{journal}{Phys. Rev. Lett.} \textbf{\bibinfo{volume}{96}},
  \bibinfo{pages}{106401} (\bibinfo{year}{2006}).

\bibitem[{\citenamefont{Liu et~al.}(2009)\citenamefont{Liu, Liu, Xu, Qi, and
  Zhang}}]{xuzhang}
\bibinfo{author}{\bibfnamefont{Q.}~\bibnamefont{Liu}},
  \bibinfo{author}{\bibfnamefont{C.-X.} \bibnamefont{Liu}},
  \bibinfo{author}{\bibfnamefont{C.}~\bibnamefont{Xu}},
  \bibinfo{author}{\bibfnamefont{X.-L.} \bibnamefont{Qi}}, \bibnamefont{and}
  \bibinfo{author}{\bibfnamefont{S.-C.} \bibnamefont{Zhang}},
  \bibinfo{journal}{Phys. Rev. Lett.} \textbf{\bibinfo{volume}{102}},
  \bibinfo{pages}{156603} (\bibinfo{year}{2009}).

\bibitem[{\citenamefont{Pesin and Balents}(2009)}]{balents2009}
\bibinfo{author}{\bibfnamefont{D.~A.} \bibnamefont{Pesin}} \bibnamefont{and}
  \bibinfo{author}{\bibfnamefont{L.}~\bibnamefont{Balents}},
  \bibinfo{journal}{arXiv:0907.2962}  (\bibinfo{year}{2009}).

\bibitem[{\citenamefont{Fu and Kane}(2008)}]{fu2008}
\bibinfo{author}{\bibfnamefont{L.}~\bibnamefont{Fu}} \bibnamefont{and}
  \bibinfo{author}{\bibfnamefont{C.~L.} \bibnamefont{Kane}},
  \bibinfo{journal}{Phys. Rev. Lett.} \textbf{\bibinfo{volume}{100}},
  \bibinfo{pages}{096407} (\bibinfo{year}{2008}).

\bibitem[{\citenamefont{Hsieh et~al.}(2009{\natexlab{b}})\citenamefont{Hsieh,
  Xia, Qian, Wray, Di, Meier, Patthey, Osterwalder, Fedorov, Lin
  et~al.}}]{hasan2009}
\bibinfo{author}{\bibfnamefont{D.}~\bibnamefont{Hsieh}},
  \bibinfo{author}{\bibfnamefont{Y.}~\bibnamefont{Xia}},
  \bibinfo{author}{\bibfnamefont{D.}~\bibnamefont{Qian}},
  \bibinfo{author}{\bibfnamefont{L.}~\bibnamefont{Wray}},
  \bibinfo{author}{\bibfnamefont{J.~H.} \bibnamefont{Di}},
  \bibinfo{author}{\bibfnamefont{F.}~\bibnamefont{Meier}},
  \bibinfo{author}{\bibfnamefont{L.}~\bibnamefont{Patthey}},
  \bibinfo{author}{\bibfnamefont{J.}~\bibnamefont{Osterwalder}},
  \bibinfo{author}{\bibfnamefont{A.}~\bibnamefont{Fedorov}},
  \bibinfo{author}{\bibfnamefont{H.}~\bibnamefont{Lin}}, \bibnamefont{et~al.},
  \bibinfo{journal}{arXiv:0904.1260}  (\bibinfo{year}{2009}{\natexlab{b}}).

\bibitem[{\citenamefont{Xu}(2009)}]{xucontinuous}
\bibinfo{author}{\bibfnamefont{C.}~\bibnamefont{Xu}},
  \bibinfo{journal}{arXiv:0909.2647}  (\bibinfo{year}{2009}).

\bibitem[{\citenamefont{Gross and Neveu}(1974)}]{gross1974}
\bibinfo{author}{\bibfnamefont{D.}~\bibnamefont{Gross}} \bibnamefont{and}
  \bibinfo{author}{\bibfnamefont{A.}~\bibnamefont{Neveu}},
  \bibinfo{journal}{Phys. Rev. D} \textbf{\bibinfo{volume}{10}},
  \bibinfo{pages}{3235} (\bibinfo{year}{1974}).

\bibitem[{\citenamefont{Wilson}(1973)}]{wilson1973}
\bibinfo{author}{\bibfnamefont{K.}~\bibnamefont{Wilson}},
  \bibinfo{journal}{Phys. Rev. D} \textbf{\bibinfo{volume}{7}},
  \bibinfo{pages}{2911} (\bibinfo{year}{1973}).

\bibitem[{\citenamefont{Hasenbusch et~al.}(1999)\citenamefont{Hasenbusch, Pinn,
  and Vinti}}]{hasenbusch1998}
\bibinfo{author}{\bibfnamefont{M.}~\bibnamefont{Hasenbusch}},
  \bibinfo{author}{\bibfnamefont{K.}~\bibnamefont{Pinn}}, \bibnamefont{and}
  \bibinfo{author}{\bibfnamefont{S.}~\bibnamefont{Vinti}},
  \bibinfo{journal}{Phys. Rev. B} \textbf{\bibinfo{volume}{59}},
  \bibinfo{pages}{11471} (\bibinfo{year}{1999}).

\bibitem[{\citenamefont{Zinn-Zustin}(1991)}]{zustin1991}
\bibinfo{author}{\bibfnamefont{J.}~\bibnamefont{Zinn-Zustin}},
  \bibinfo{journal}{Nucl. Phys. B} \textbf{\bibinfo{volume}{367}},
  \bibinfo{pages}{105} (\bibinfo{year}{1991}).

\bibitem[{\citenamefont{Karkkainen et~al.}(1994)\citenamefont{Karkkainen,
  Lacaze, Lacock, and Petersson}}]{petersson1994}
\bibinfo{author}{\bibfnamefont{L.}~\bibnamefont{Karkkainen}},
  \bibinfo{author}{\bibfnamefont{R.}~\bibnamefont{Lacaze}},
  \bibinfo{author}{\bibfnamefont{P.}~\bibnamefont{Lacock}}, \bibnamefont{and}
  \bibinfo{author}{\bibfnamefont{B.}~\bibnamefont{Petersson}},
  \bibinfo{journal}{Nucl. Phys. B} \textbf{\bibinfo{volume}{415}},
  \bibinfo{pages}{781} (\bibinfo{year}{1994}).

\bibitem[{\citenamefont{Gracey}(1991)}]{gracey1991}
\bibinfo{author}{\bibfnamefont{J.~A.} \bibnamefont{Gracey}},
  \bibinfo{journal}{Int. J. Mod. Phys. A} \textbf{\bibinfo{volume}{6}},
  \bibinfo{pages}{395} (\bibinfo{year}{1991}).

\bibitem[{\citenamefont{Gracey}(1992)}]{gracey1992}
\bibinfo{author}{\bibfnamefont{J.~A.} \bibnamefont{Gracey}},
  \bibinfo{journal}{Phys. Lett. B} \textbf{\bibinfo{volume}{297}},
  \bibinfo{pages}{293} (\bibinfo{year}{1992}).

\bibitem[{\citenamefont{Hertz}(1976)}]{hertz1976}
\bibinfo{author}{\bibfnamefont{J.~A.} \bibnamefont{Hertz}},
  \bibinfo{journal}{Phys. Rev. B} \textbf{\bibinfo{volume}{14}},
  \bibinfo{pages}{1165} (\bibinfo{year}{1976}).

\bibitem[{\citenamefont{Polchinski}(1994)}]{polchinksi1994}
\bibinfo{author}{\bibfnamefont{J.}~\bibnamefont{Polchinski}},
  \bibinfo{journal}{Nucl. Phys. B} \textbf{\bibinfo{volume}{422}},
  \bibinfo{pages}{617} (\bibinfo{year}{1994}).

\bibitem[{\citenamefont{Lee}(2009)}]{sslee2009}
\bibinfo{author}{\bibfnamefont{S.-S.} \bibnamefont{Lee}},
  \bibinfo{journal}{arXiv: 0905.4532}  (\bibinfo{year}{2009}).

\bibitem[{\citenamefont{Zhang et~al.}(2009{\natexlab{b}})\citenamefont{Zhang,
  Liu, Qi, Deng, Dai, Zhang, and Fang}}]{liu2009}
\bibinfo{author}{\bibfnamefont{H.-J.} \bibnamefont{Zhang}},
  \bibinfo{author}{\bibfnamefont{C.-X.} \bibnamefont{Liu}},
  \bibinfo{author}{\bibfnamefont{X.-L.} \bibnamefont{Qi}},
  \bibinfo{author}{\bibfnamefont{X.-Y.} \bibnamefont{Deng}},
  \bibinfo{author}{\bibfnamefont{X.}~\bibnamefont{Dai}},
  \bibinfo{author}{\bibfnamefont{S.-C.} \bibnamefont{Zhang}}, \bibnamefont{and}
  \bibinfo{author}{\bibfnamefont{Z.}~\bibnamefont{Fang}},
  \bibinfo{journal}{arXiv:0901.2762}  (\bibinfo{year}{2009}{\natexlab{b}}).

\bibitem[{\citenamefont{Fu}(2009)}]{fu2009}
\bibinfo{author}{\bibfnamefont{L.}~\bibnamefont{Fu}},
  \bibinfo{journal}{arXiv:0908.1418}  (\bibinfo{year}{2009}).

\bibitem[{\citenamefont{Rashba}(1960)}]{rashba1960}
\bibinfo{author}{\bibfnamefont{E.~I.} \bibnamefont{Rashba}},
  \bibinfo{journal}{Sov. Phys. Solid State} \textbf{\bibinfo{volume}{2}},
  \bibinfo{pages}{1106} (\bibinfo{year}{1960}).

\bibitem[{\citenamefont{Bychkov and Rashba}(1984)}]{rashba1984}
\bibinfo{author}{\bibfnamefont{Y.~A.} \bibnamefont{Bychkov}} \bibnamefont{and}
  \bibinfo{author}{\bibfnamefont{E.~I.} \bibnamefont{Rashba}},
  \bibinfo{journal}{J. Phys. C} \textbf{\bibinfo{volume}{17}},
  \bibinfo{pages}{6039} (\bibinfo{year}{1984}).

\end{thebibliography}
\end{document}